\documentclass[prb,a4,twocolumn,showpacs,amsmath,amssymb]{revtex4-1}
\usepackage{graphicx}
\usepackage{graphicx,subfigure,epsfig}
\usepackage{color}
\usepackage{soul}
\usepackage{float}
\usepackage{amsmath}
\usepackage{amssymb}
\usepackage{ifthen}
\usepackage{alltt}
\usepackage{fancyhdr}
\usepackage{verbatim}
\usepackage{moreverb}
\usepackage{colortbl}
\usepackage{multirow}
\usepackage{bigstrut}
\usepackage{wrapfig}
\usepackage{enumerate}
\usepackage{rotating}
\usepackage{epstopdf}

\begin{document}

\title{Topological phase transition in wire medium enables high Purcell factor at infrared frequencies}
\author{M.~S.~Mirmoosa$^1$, S.~Yu.~Kosulnikov$^{1,2}$ and C.~R.~Simovski$^{1,2}$}
\affiliation{$^1$Department of Radio Science and Engineering, School of Electrical Engineering, Aalto University, P.O.~Box 13000, FI-00076 AALTO, Finland\\
$^2$Department of Nanophotonics and Metamaterials, ITMO University, 197043, St.~Petersburg, Russia}
\date{\today }


\begin{abstract}
In this paper, we study topological phase transition in a wire medium operating at infrared frequencies. This transition occurs in the reciprocal space
between the indefinite (open-surface) regime of the metamaterial to its dielectric (closed-surface) regime.
Due to the spatial dispersion inherent to wire medium, a hybrid regime turns out to be possible at the transition frequency. Both such surfaces exist at the same frequency and touch one another.
At this frequency, all values of the axial wavevector correspond to propagating spatial harmonics. The implication of this
regime is the overwhelming radiation enhancement. We numerically investigated the gain in radiated power for a sub-wavelength dipole source submerged into such the medium.
In contrast to all previous works, this gain (called the Purcell factor) turns out to be higher for an axial dipole than for a transversal one.
\end{abstract}

\maketitle


\section{Introduction}

Electromagnetic metamaterials are artificial media structured on the sub-wavelength level which exhibit extraordinary electromagnetic features and functionalities that are not found in nature. Most popular examples of these unusual functionalities can be found in works \cite{veselago,pendry,leonhardt}. Many metamaterials possess unusual dispersion properties. For regular metamaterials (lattices) their dispersion properties are described in terms of dispersion surfaces, also called Fresnel's wave or iso-frequency surfaces. A dispersion surface defines wavevectors for all propagation directions at the given frequency in the reciprocal space of the lattice. One of the striking examples of metamaterials are hyperbolic ones called this way because their dispersion surface is hyperboloid. Such metamaterials are exploited in a variety of applications ranging from sub-wavelength imaging to radiative heat transfer \cite{kivshar}. Indeed, a non-magnetic hyperbolic metamaterial is most popular in optics, where magnetic properties are very difficult to realize. Optical hyperbolic metamaterials are anisotropic media whose principal components (more exactly -- real parts) of the effective permittivity tensor have different signs. Because of the indefinite sign of the permittivity tensor trace such media are also called indefinite \cite{smith}. Most expanded are indefinite (hyperbolic) media of uniaxial type. The different signs of the axial and transverse components of the permittivity tensor imply the hyperboloid of revolution in the reciprocal space. Since the hyperboloid is infinitely extended, the photonic density of states in hyperbolic metamaterials is very high \cite{kivshar}. It results in the enhancement of the radiated power for a dipole radiator embedded in such a medium \cite{kivshar,jacob,belov}. Usually, this enhancement of radiation is used for optical sensing and locating the fluorescent nanoobjects. The gain in the radiated power at the frequency of spontaneous emission by the fluorescent object implies the increase of the decay rate. This increase is called Purcell factor and may attain two orders of magnitude for emitters embedded into a hyperbolic metamaterial \cite{jacob,belov,yan}. Initially, the Purcell effect \cite{purcell1,purcell2} was treated as an increase of the spontaneous decay rate of an emitter located in a resonant cavity.
Recently, this concept was generalized to the modification of the decay rate in presence of any environment different from free space \cite{purcell}.

Uniaxial hyperbolic metamaterials operating in the optical range are usually implemented as a stack of bilayers. One layer of the unit cell is a plasmonic metal, another layer is a transparent material (dielectric or semiconductor). The thickness of the plasmonic layer is smaller than the skin-depth that makes the unit cell penetrable for propagating waves. The total thickness of the unit cell is much smaller than the wavelength in the transparent material. Another known implementation of hyperbolic metamaterial is a regular optically dense  array of parallel metal wires called wire medium (see e.g. in \cite{kivshar,constantin}). The homogenization models known for both kinds of hyperbolic metamaterials (see in \cite{kivshar,constantin,chebykin,mario}) point out to the possibility of a sharp qualitative change of their dispersion regime versus frequency. When the frequency varies one of two principal components of the permittivity tensor ($\varepsilon_{\parallel}$ or $\varepsilon_{\perp}$) may keep its sign whereas the sign of the other component changes. Then the hyperboloid in the reciprocal space transforms into a closed surface, e.g. an ellipsoid. This transition is similar to the known Lifshitz topological transitions in metals \cite{lifshitz} (when the Fermi surfaces change their topology similarly). Therefore, this jump of the dispersion properties was entitled \textquotedblleft{topological phase transition}\textquotedblright in Ref.~\cite{krishnamoorthy}. Topological transition may additionally enhance the spatial spectrum of propagating eigenmodes of the medium and, consequently, the gain in the spontaneous decay rate/radiated power of an emitter located in this medium. In fact, the hyperboloid though infinitely extended implies the cut-off for the axial component of the wavevector. This cut-off at a given frequency $\omega$ is equal to $k_0\varepsilon_{\perp}$, where $k_0=\omega/c$ is the free-space wave number. The spatial harmonics whose axial wavevector is below cut-off do not propagate.

In Ref.~\cite{krishnamoorthy}, a constitutive bilayer of the hyperbolic metamaterial is formed by titanium oxide and silver nanolayers. At the topological transition frequency, $\varepsilon_{\perp}$ of this metamaterial tends to zero in the framework of the homogenization model (neglecting optical losses). Then the distance between two branches of the hyperbola squeezes and the cut-off vanishes. The spatial spectrum at this frequency is not only infinitely extent, it is unbounded. This obviously grants the maximal possible density of photonic states. Therefore, the Purcell factor calculated in Ref.~\cite{krishnamoorthy} achieved two orders of magnitude even for a quantum dot located on the surface of the hyperbolic metamaterial. Of course, optical losses
decrease this effect, however, in work \cite{yang} it was experimentally shown (in this case for a stack of silver-germanium bilayers) that even the radiative Purcell factor -- the gain in power radiated to free space one achieves one order of magnitude for emitters located on the surface. Further, in Ref.~\cite{gomez} one presented a general theoretical study of the phenomenon of unbounded spatial spectrum where the possible magnetic response (effective permeability tensor) was also involved.

We think that the strong Purcell effect obtained at the topological transition of a stacked hyperbolic metamaterial does not close the chapter. We can study the other realization of a hyperbolic metamaterial, wire medium which can be fabricated with free-standing parts \cite{molesky}, alternatively nanowires can partially stay in a liquid host \cite{kabashin}. This allows quantum emitters to easily emerge in the bulk. Moreover, wire medium has its own advantages in nanofabrication. Fraction ratio and parallelism of nanowires are important, but the strict periodicity is not crucial for material parameters of the wire medium, whereas the stacked implementation of hyperbolic metamaterial is very demanding to the flatness and periodicity of constitutive nanolayers.

In the available literature topological transitions in wire media, to our knowledge, has not been revealed, yet. This is not surprising because most popular wire media are implemented of plasmonic (Ag or Au) nanowires. For plasmonic wire media with practical design parameters topological transitions, to our estimations, occur at frequencies where the homogenization model loses its validity. Since the homogenization effective-medium model is analytical, it represents the best tool for searching the needed regime. Therefore, we inspected different materials with the use of the effective-medium model in order to find the needed regime. The finding was validated via the exact calculation of the Purcell factor.

In this paper, we report the topological transition in a wire medium whose nanowires are prepared of LiTaO$_3$. It occurs in the mid IR range, where the array period is much smaller than the wavelength and the homogenization model keeps adequate. Both hyperbolic and elliptic dispersion regimes are combined at the transition frequency, and the corresponding dispersion branches are connected that offers the unbounded spatial spectrum. Though optical losses in such the wire medium are substantial they are not overcritical and the unbounded spatial spectrum manifests itself in the high Purcell factor. This \textquotedblleft{transitional}\textquotedblright Purcell factor should be, evidently, higher than that obtained in Ref.~\cite{poddubny} for a wire medium operating in the \textquotedblleft{standard}\textquotedblright hyperbolic regime (with a cut-off). Since the spatial spectrum is unbounded over the axial wavevector the axial dipole should generate more radiation than the transverse one. This unusual expectation diverges from all previously known results. The Purcell factor of \textquotedblleft{standard}\textquotedblright (hyperbolic) wire media attains two orders of magnitude for a transverse dipole, whereas
for an axial dipole it is smaller \cite{kivshar,belov,poddubny,vorobev}. In the transition regime the Purcell factor of a transverse dipole keeps the same as in the hyperbolic regime, whereas for an axial dipole it exceeds two orders of magnitude.

Our paper is organized as follows: in Sec.~\ref{sec:theory}, we introduce the approach how to achieve the unbounded spatial spectrum in a wire medium. Section~\ref{sec:results} presents the numerical results for the Purcell factor and gives some discussions on their interpretation. Finally, Sec.~\ref{sec:conclusion} concludes the paper.


\section{Theory}
\label{sec:theory}
\begin{figure}[h!]\centering
\includegraphics[width=8cm]{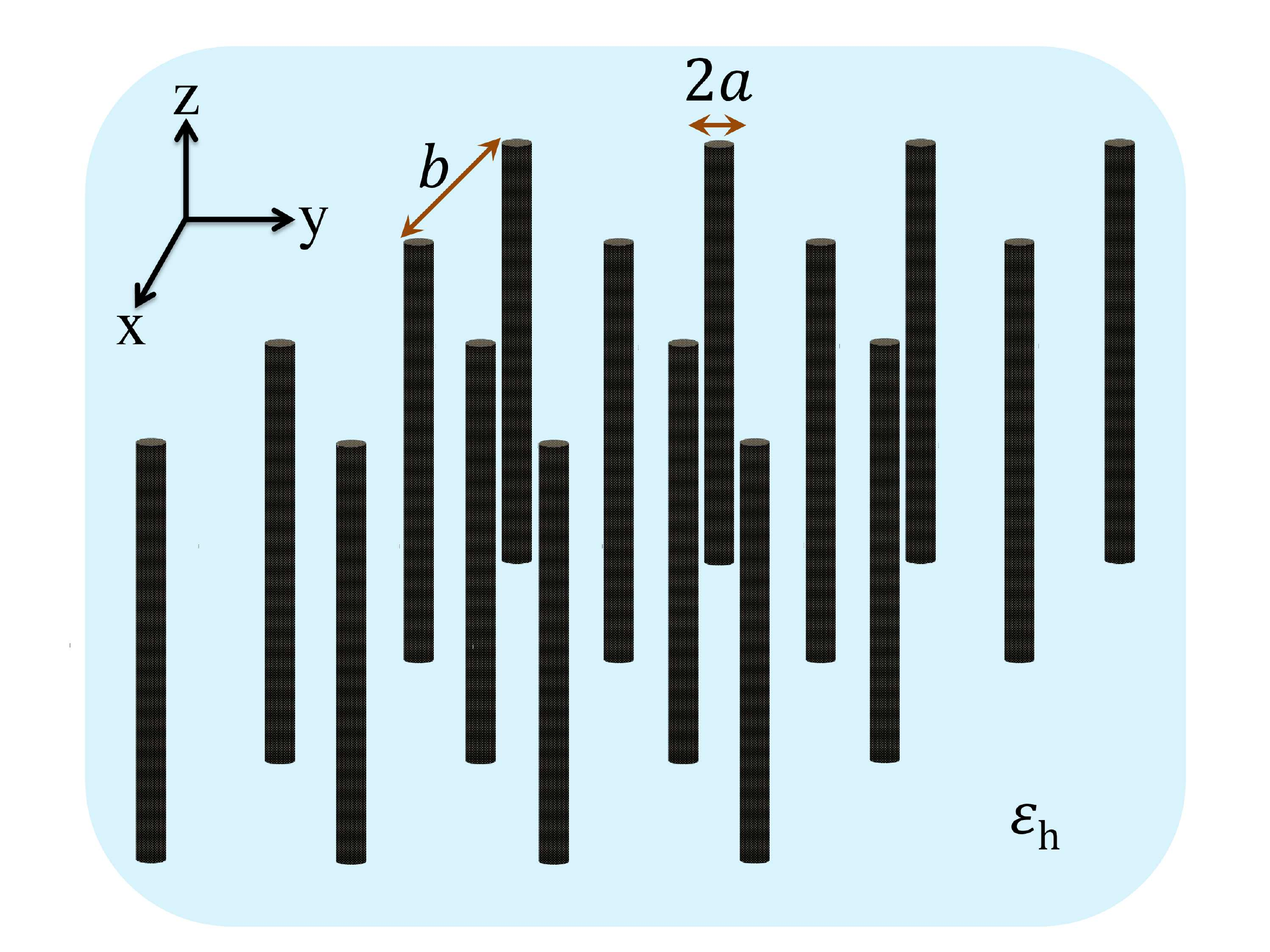}
\caption{Long parallel wires which are set in a square lattice. $b$ and $a$ are the lattice constant and the radius of a dielectric rod, respectively, and $\varepsilon_{\rm{h}}$ represents the host medium.}
\label{fig:wm}
\end{figure}

Electromagnetic properties of a general wire medium shown in Fig.~\ref{fig:wm} is described through a homogeneous effective permittivity tensor as
\begin{equation}
\overline{\overline{\epsilon}}=\left(\begin{array}{ccc}
\varepsilon_{\perp} & 0 & 0 \\
0 &\varepsilon_{\perp} & 0 \\
0& 0 & \varepsilon_{\parallel} \end{array} \right),
\end{equation}
where in case of thin and negative-dielectric constant wires, the transverse ($\varepsilon_\perp$) and axial ($\varepsilon_\parallel$) components of the effective tensor are given by \cite{mario}
\begin{equation}
\begin{split}
&\varepsilon_{\perp}=\varepsilon_{\rm{h}}+\displaystyle\frac{2\varepsilon_{\rm{h}}}{\displaystyle\frac{\varepsilon_{\rm{r}}+\varepsilon_{\rm{h}}}{f_{\rm{v}}(\varepsilon_{\rm{r}}-\varepsilon_{\rm{h}})}-1},\\
&\varepsilon_{\parallel}=\varepsilon_{\rm{h}}+\displaystyle\frac{\varepsilon_{\rm{h}}}{\displaystyle\frac{\varepsilon_{\rm{h}}}{f_{\rm{v}}(\varepsilon_{\rm{r}}-\varepsilon_{\rm{h}})}-\displaystyle\frac{k_0^2-\beta^2}{k_{\rm{p}}^2}}.
\end{split}
\label{eq:permittivity}
\end{equation}
Here, $\varepsilon_{\rm{h}}$ and $\varepsilon_{\rm{r}}$ are the permittivity of the host medium and the wires, respectively, and $f_{\rm{v}}$ denotes the volume fraction. In this paper, we assume that the host medium is air, i.e. $\varepsilon_{\rm{h}}=1$. In addition, $k_0$ is the free-space wave number, $\beta$ is the axial component of the wavevector in the wire medium, and $k_{\rm{p}}$ is called the plasma wave-number:
\begin{equation}
k_{\rm{p}}=\displaystyle\frac{1}{b}\displaystyle\sqrt{\displaystyle\frac{2\pi}{0.5275+\ln{\left(\displaystyle\frac{b}{2\pi a}\right)}}}.
\label{eq:plasma}
\end{equation}
Since the wire medium is a uniaxial, two modes exist in it. The first mode is ordinary transverse-electric (TE) wave, and the second one is extraordinary transverse-magnetic (TM) wave. In this work, we are interested in the extraordinary wave because the hyperbolic dispersion corresponds only to this mode. Solving Maxwell's equations, the dispersion relation of extraordinary wave can be expressed as \cite{marcuvitz}
\begin{equation}
\displaystyle\frac{q^2}{\varepsilon_\parallel(\beta)}+\displaystyle\frac{\beta^2}{\varepsilon_\perp}=k_0^2,
\label{eq:dispersion}
\end{equation}
where $q$ is the transverse component of the wavevector. Here, $\varepsilon_\parallel$ is a function of $\beta$.

For simplicity, let us assume that the wires are lossless.
If their volume fraction is small enough and the relative dielectric constant of the wire material is not very close to $-1$, the transverse component $\varepsilon_\perp$ is positive and approximately equals to unity (in the general case -- to that of the host dielectric material). Equation~\ref{eq:dispersion} indicates that if $\varepsilon_\parallel<0$, a hyperbolic dispersion arises, and if $\varepsilon_\parallel>0$, an elliptical dispersion arises. Since hyperbolic dispersion implies propagation of waves with high values of both axial and transverse wavevectors, the evanescent waves usually produced by a subwavelength dipole source are converted into propagating waves. This results in the high Purcell factor, as it was already mentioned in the introduction.

As it is seen from Eq.~\ref{eq:dispersion}, there is a cut-off for axial component of the wavevector: $\beta\ge k_0\sqrt{\varepsilon_\perp}$.
To increase the Purcell factor we need to eliminate this cut-off. For $\beta<k_0\sqrt{\varepsilon_\perp}$, based on Eq.~\ref{eq:dispersion}, the propagation becomes possible if $\varepsilon_{\perp}>0$ (elliptical dispersion). We need to unify both elliptical and hyperbolic dispersions at the same frequency so that these branches touch each other at the transition point $\beta=k_0\sqrt{\varepsilon_\perp}$. It is possible because the axial component of permittivity in Eq.~\ref{eq:dispersion} depends on the wavevector.

If we substitute Eq.~\ref{eq:permittivity} in Eq.~\ref{eq:dispersion}, the transversal wavevector can be expressed as
\begin{equation}
q^2=\displaystyle\frac{1}{\varepsilon_\perp}\displaystyle\left[\beta^2-R_1\right]\displaystyle\left[\beta^2-R_2\right]\displaystyle\frac{f_{\rm{v}}(1-\varepsilon_{\rm{r}})}{k_{\rm{p}}^2-f_{\rm{v}}(\varepsilon_{\rm{r}}-1)(k_0^2-\beta^2)},
\label{eq:wave vector}
\end{equation}
where
\begin{equation}
\begin{split}
&R_1=\displaystyle\frac{f_{\rm{v}}(\varepsilon_{\rm{r}}-1)k_0^2-(1+f_{\rm{v}}(\varepsilon_{\rm{r}}-1))k_{\rm{p}}^2}{f_{\rm{v}}(\varepsilon_{\rm{r}}-1)},\\
&R_2=k_0^2\varepsilon_\perp.
\end{split}
\label{eq:roots}
\end{equation}
If the roots, $R_1$ and $R_2$ are equivalent and $f_{\rm{v}}(1-\varepsilon_{\rm{r}})>0$, the value $q^2$ is always positive which results in propagating mode for any $\beta$.  The cut-off
of spatial spectrum is removed at this frequency, and we have
\begin{equation}
f_{\rm{v}}(1-\varepsilon_{\rm{r}})=\displaystyle\frac{k_{\rm{p}}^2}{k_{\rm{p}}^2+(\varepsilon_\perp-1)k_0^2}.
\label{condition}
\end{equation}
Since the plasma wave number is very large compared to $\sqrt{\varepsilon_\perp-1}k_0$ ($k_{\rm{p}}^2\gg(\varepsilon_\perp-1)k_0^2$), the needed permittivity of the wire material can be approximated as
\begin{equation}
\varepsilon_{\rm{r}}=1-\displaystyle\frac{1}{f_{\rm{v}}}.
\label{eq:condition}
\end{equation}
Condition expressed in Eqs.~\ref{condition} and \ref{eq:condition} results in $\varepsilon_\parallel=0$ at the point $\beta=k_0\sqrt{\varepsilon_\perp}$ in the iso-frequency surface. Based on Eq.~\ref{eq:permittivity}, it can be easily shown that for $\beta<k_0\sqrt{\varepsilon_\perp}$, $\varepsilon_\parallel$ is positive (corresponding to nearly elliptical dispersion), and for $\beta>k_0\sqrt{\varepsilon_\perp}$, $\varepsilon_\parallel$ is smaller than zero (corresponding to nearly hyperbolic dispersion). This topological transition at the desired frequency removes the cut-off in the axial wavevector. Since the permittivity tensor depends on the wavevector, we do not obtain exactly hyperboloidal and ellipsoidal parts of the dispersion surface. Therefore, instead of hyperboloid and ellipsoid, we prefer to use more exact terms: open surface and closed surface, respectively.

\begin{figure}[t!]\centering
\subfigure[]{\includegraphics[width=9.15cm]{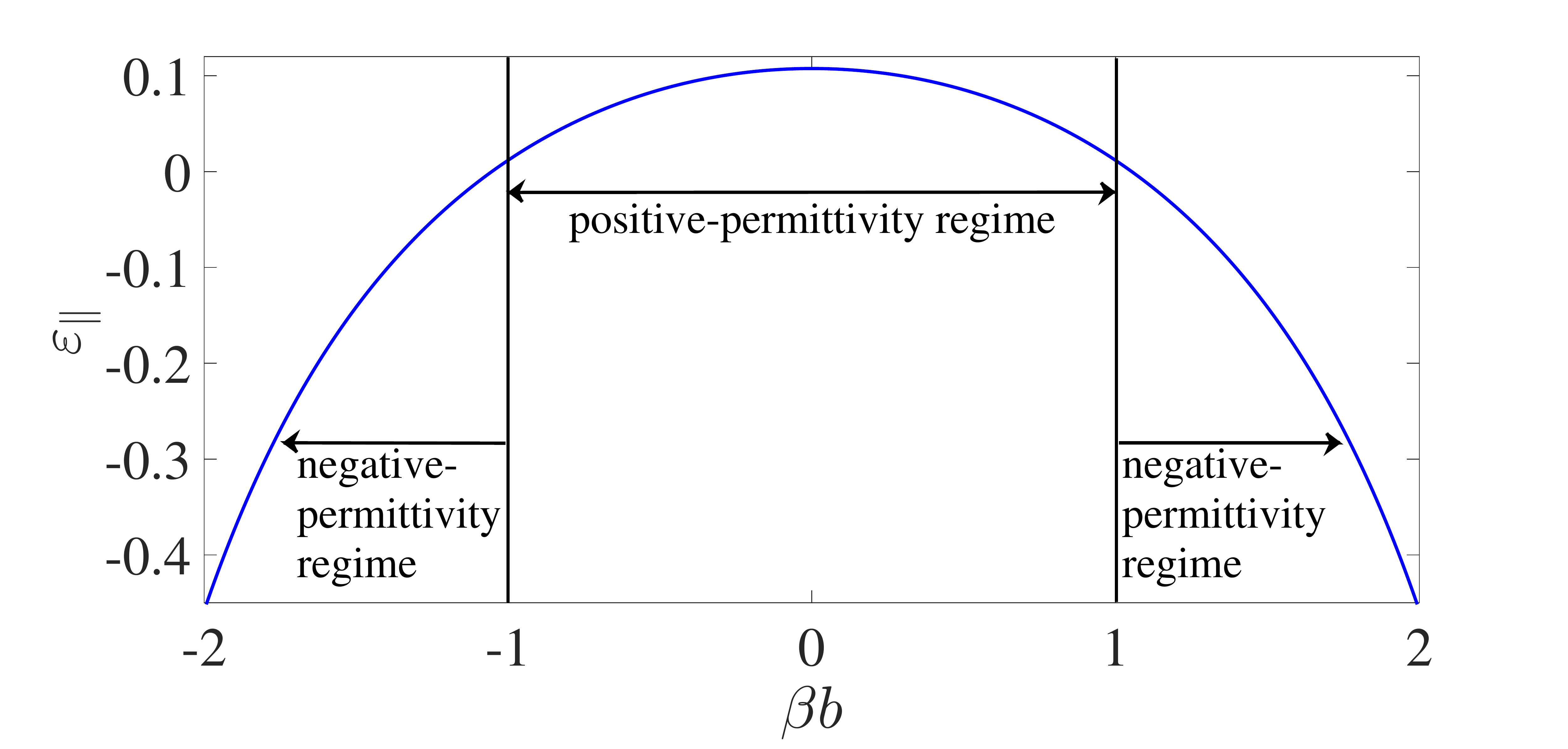}
\label{fig:perm}}
\subfigure[]{\includegraphics[width=9.15cm]{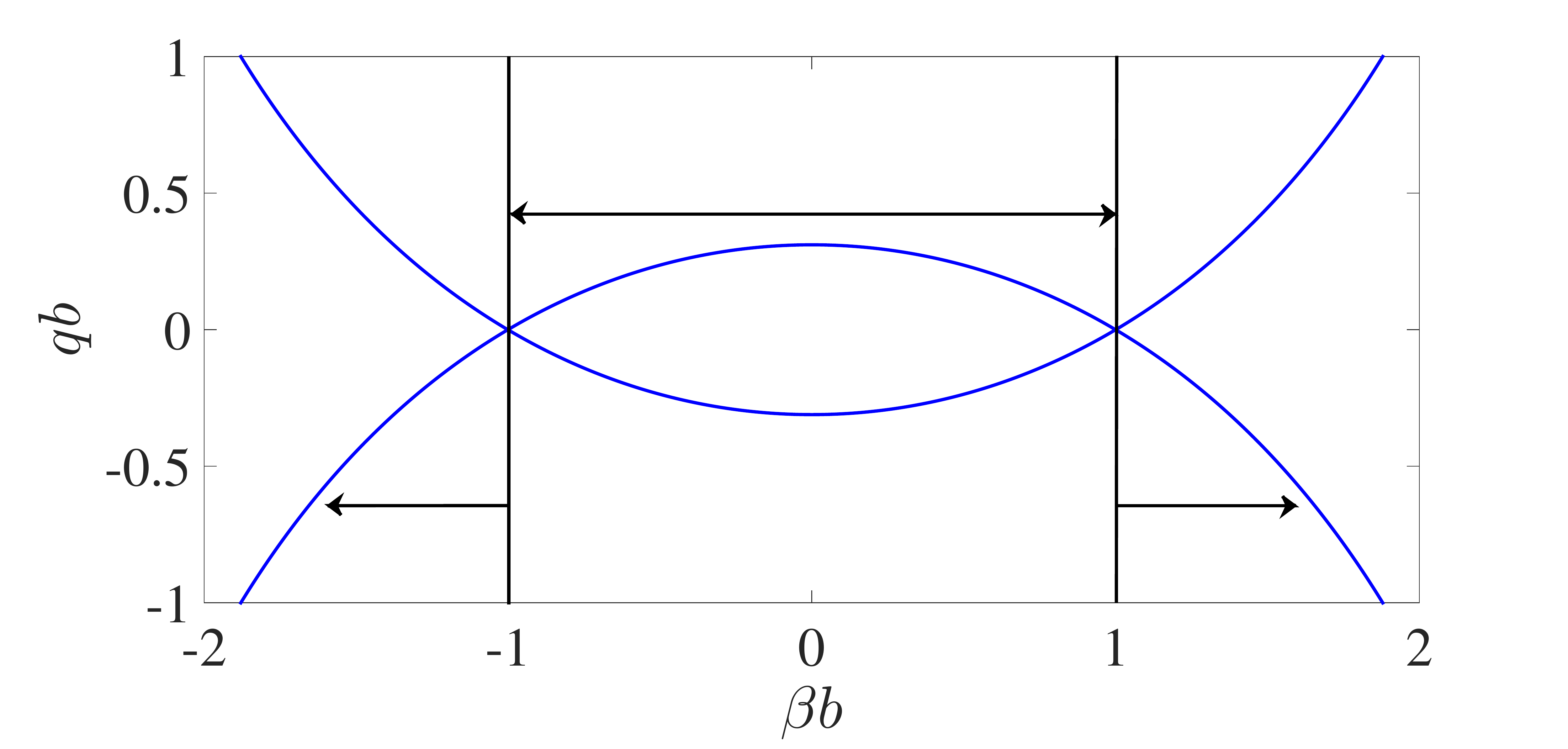}
\label{fig:iso}}
\caption{(a)--Axial component of effective permittivity with respect to normalized wavevector and (b)--corresponding iso-frequency contour (cross section of the dispersion surface). Calculations are related to the following values: $f_{\rm{v}}=0.05$, and $\lambda/b=6.6$ where $\lambda$ is the free-space wavelength.}
\end{figure}

\begin{figure}[t!]\centering
\subfigure[]{\includegraphics[width=9.15cm]{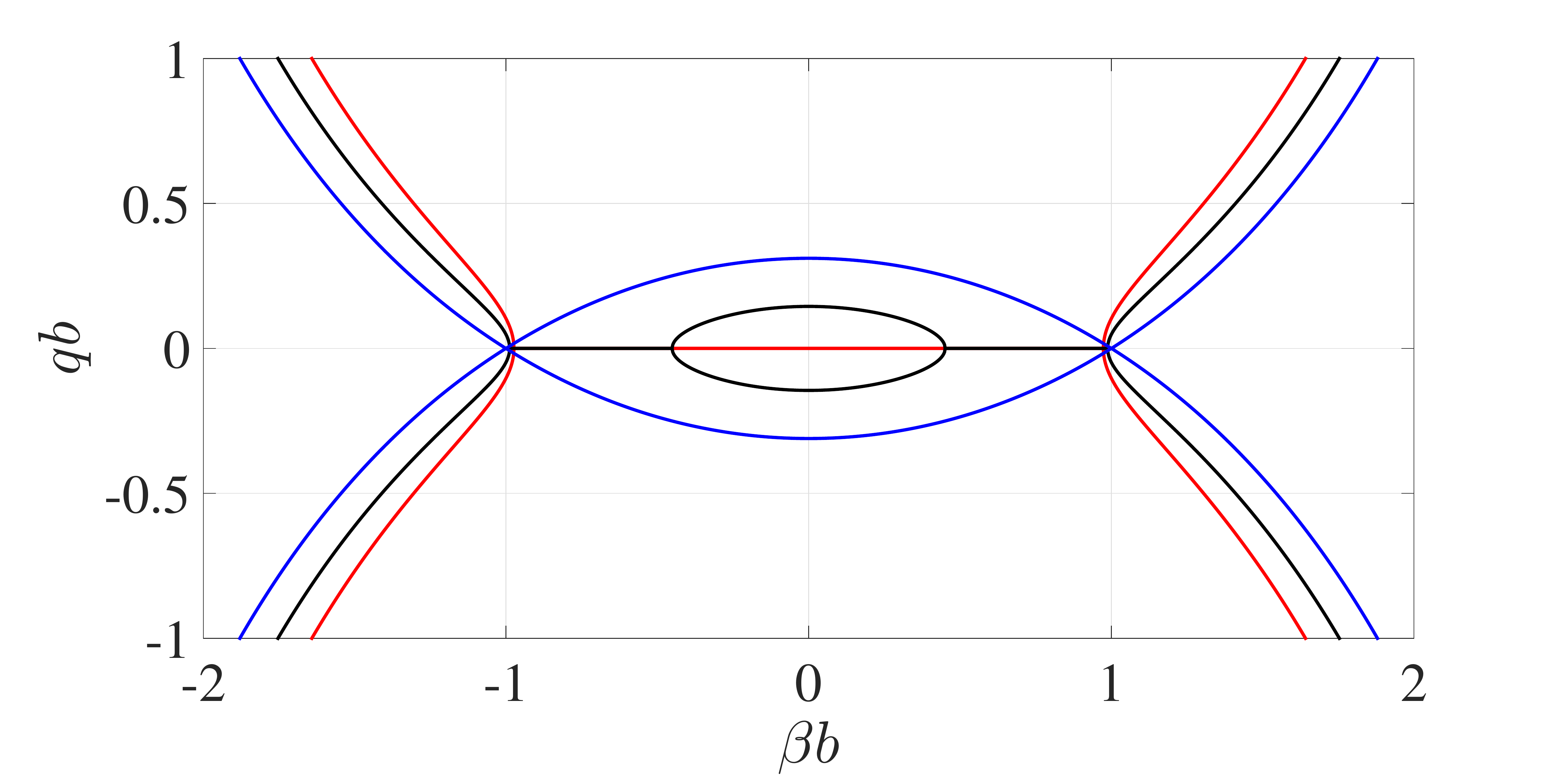}}
\subfigure[]{\includegraphics[width=9.15cm]{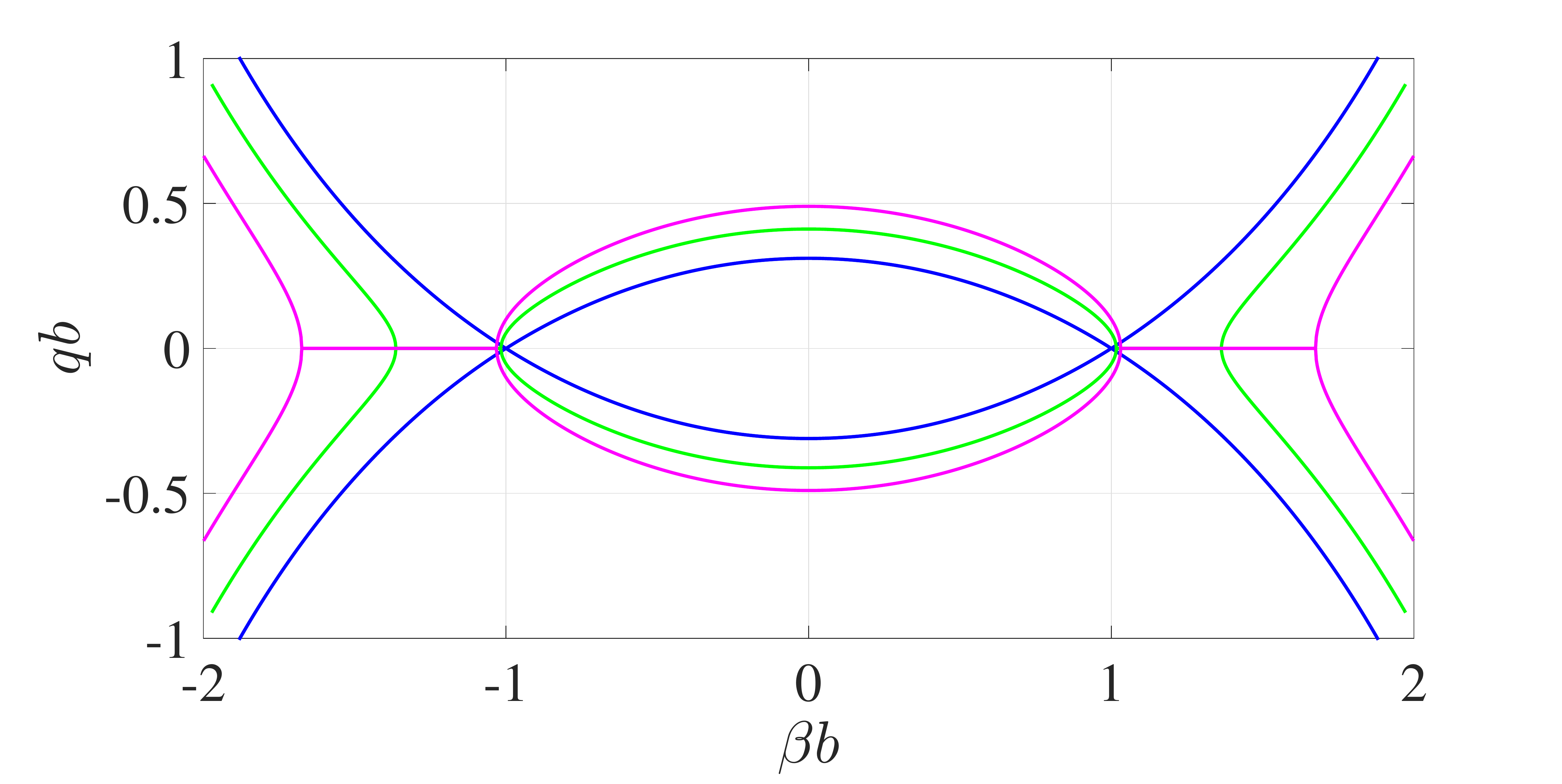}}
\caption{Iso-frequency contours for different values of the relative dielectric constant of the wires. Calculations are related to the following values: $f_{\rm{v}}=0.05$, and $\lambda/b=6.6$. The curves based on their colors correspond to red--$\varepsilon_{\rm{r}}=-23$, black--$\varepsilon_{\rm{r}}=-20.75$, blue--$\varepsilon_{\rm{r}}=-19$, green--$\varepsilon_{\rm{r}}=-17$ and magneta--$\varepsilon_{\rm{r}}=-15.35$.}
\label{fig:isofre}
\end{figure}

Fig.~\ref{fig:perm} shows the axial component of the effective permittivity versus the normalized axial wavevector at the optimized wavelength in which the topological transition happens. In this example the fraction of wires $f_{\rm{v}}=0.05$ that makes the relative dielectric constant of the wire material equal to $\varepsilon_{\rm{r}}=-19$ in according to Eq.~\ref{eq:condition}. The figure has specified the closed-surface (positive-permittivity) and the open-surface (negative-permittivity) regimes. The transition happens at $\beta b=\pm1$ where $\varepsilon_\parallel(\beta)=0$. The corresponding iso-frequency curve is shown in Fig.~\ref{fig:iso}. Different iso-frequency surfaces for several values of relative dielectric constant of the wires are shown in Fig.~\ref{fig:isofre}. It is easily seen that the optimal regime corresponds to the blue solid curve representing $\varepsilon_{\rm{r}}=-19$ when there is no spatial-frequency cut-off. When $\varepsilon_{\rm{r}}=-23$, the closed surface does not exist and the cut-off is present. Reducing gradually the absolute value of the relative dielectric constant causes the positive-permittivity regime to appear. Both parts of the iso-frequency surface touch each other at the point $\varepsilon_\parallel=0$ if $\varepsilon_{\rm{r}}=-19$.

So, the topological transition in a wire medium  whose wires have relative permittivity close to $-19$ is possible in the frequency range where the non-local homogenization model \cite{mario} is applicable. This model has been recognized for its high accuracy \cite{constantin}. At the corresponding frequency, the spatial spectrum is unbounded (no cut-off as for the stacked implementation of the hyperbolic metamaterial at its topological transition \cite{krishnamoorthy}). However, this effect happens not because the distance between two hyperbolic branches squeezes as in~\cite{krishnamoorthy}. In our case both closed-surface and open-surface branches exist in the reciprocal space together and touch one another.

Notice, that such the phenomenon is, in principle, known in the theory of metamaterials with spatial dispersion. In work \cite{gorlach} one has suggested and theoretically studied a spatially dispersive metamaterial for which both hyperboloid and ellipsoid surfaces exist in the reciprocal space simultaneously (beyond a topological transition). The suggested metamaterial in \cite{gorlach} is an ultimately anisotropic strictly regular lattice of uniaxial strongly resonant inclusions. In optics such an array could be implemented from very thin and substantially long Ag or Au nanorods operating at their plasmon resonance. Varying the geometric parameters one may engineer the regime of unbounded spatial spectrum when the open and closed branches of the dispersion surface would be touching or intersecting.

However, the metamaterial \cite{gorlach} is challenging for practical realization at optical frequencies, and its experimental verification in the near future is problematic. Meanwhile,
the spatial dispersion which is needed to unify the elliptic and hyperbolic dispersion branches turns out to be possible in a wire medium. In the next section, we consider an implementation of
the wire medium of a polaritonic crystalline material. Aligned nanowires of polaritonic materials grown in a solid matrix have been reported since 2009 where the first paper on them was published  \cite{polaritonics}. Though for the instance polaritonic wire media with free-standing parts are not known, there are no basic technological limitations, and we hope that they will be reported soon. Moreover, the topological transition with needed spatial dispersion is not a prerogative of polaritonic crystals. We plan to find the needed regime for polycrystal media more popular among nanotechnologists such as SiC or  GaN which possess the needed modest negative relative permittivity in the mid IR range.


\section{Results and Discussion}
\label{sec:results}

To confirm our expectation for the Purcell factor at the topological phase transition, we performed full-wave simulations employing the CST Microwave Studio simulator. For reliability, some simulation results were also reproduced using the HFSS software. The sample of wire medium in our simulations is cubic, and the volume fraction of lithium tantalate is equal to $f_{\rm{v}}=0.0804$. This value may correspond, for example to the wire radius $a=32$ nm and period $b=200$ nm. Based on Eq.~\ref{eq:condition}, the relative dielectric constant of the wires should have ${\rm Re}(\varepsilon_{\rm{r}})= -11.4$ and $|{\rm Im}(\varepsilon_{\rm{r}})|\ll |{\rm Re}(\varepsilon_{\rm{r}})|$ that allows $\varepsilon_\parallel\approx 0$ at the point $\beta=k_0\sqrt{\varepsilon_\perp}$ and subsequently the transition from negative-permittivity regime (${\rm Re}(\varepsilon_\parallel)<0$) to positive-permittivity regime (${\rm Re}(\varepsilon_\parallel)>0$). Because ${\rm Re}(\varepsilon_{\rm{r}})$ is negative, the transition is in principle possible for metal nanowires within visible range.
However, this option requires either very tiny nanowires which will be challenging for fabrication or more sparse array for which the transition occurs at the wavelength beyond the domain of applicability of the homogenization model. Another option is using polaritonic materials, e.g. lithium tantalate operating in the near-IR or mid-IR ranges. The dielectric function of such materiales is provided by the Drude-Lorentz model \cite{kittel}: $\varepsilon_{\rm{r}}=\varepsilon_\infty[1+(\omega_{\rm{L}}^2-\omega_{\rm{T}}^2)/(\omega_{\rm{T}}^2-\omega^2+j\omega\gamma)]$ where for
LiTaO$_3$ we have $\omega_{\rm{T}}/2\pi=26.7$ THz, $\omega_{\rm{L}}/2\pi=46.9$ THz, $\gamma/2\pi=0.94$ THz, and $\varepsilon_\infty=13.4$ \cite{schall,crimmins}. The value $\varepsilon_{\rm{r}}=-11.4$ corresponds to the permittivity of lithium tantalate at about $f=39$ THz.

The emitter is chosen to be a sub-wavelength electric dipole. To have studied different orientations and locations of this dipole inside the wire medium sample. Here, we show the results for both longitudinal (dipole parallel to nanowires and optical axis) and transversal (dipole perpendicular to optical axis) orientations and compare the results. In both cases the dipoles are located symmetrically in the gap between nanowires at the center of the wire medium sample, as it is shown in Fig.~\ref{fig:vertical}.

We calculated two types of the Purcell factor. One is called full Purcell factor ($FP_{\rm{F}}$) which shows the total enhancement of the radiated power of the dipole. It can be found through the
real part of the input impedance of the lossless Hertzian dipole which is equal to its radiation resistance. Keeping the same current in the dipole in the presence of the metamaterial sample and
in its absence the ratio of radiation resistances delivers the gain in the radiated power $FP_{\rm{F}}$. The second type of the Purcell factor is called radiative Purcell factor ($RP_{\rm{F}}$). It determines the enhancement of the power radiated by the same dipole moment into free space. $RP_{\rm{F}}$ is calculated via the power flux integrated in the far zone in the presence of the sample and in its absence. An important check of the reliability of our simulations was done nullifying the optical losses in lithium tantalate \textquotedblleft{by hands}\textquotedblright. Then the absorption of radiated power in the sample, i.e. the non-radiative part of the full Purcell factor vanishes and $FP_{\rm{F}}=RP_{\rm{F}}$.

\subsection{Longitudinal dipole emitter}
\begin{figure}[h!]\centering
\includegraphics[width=8cm]{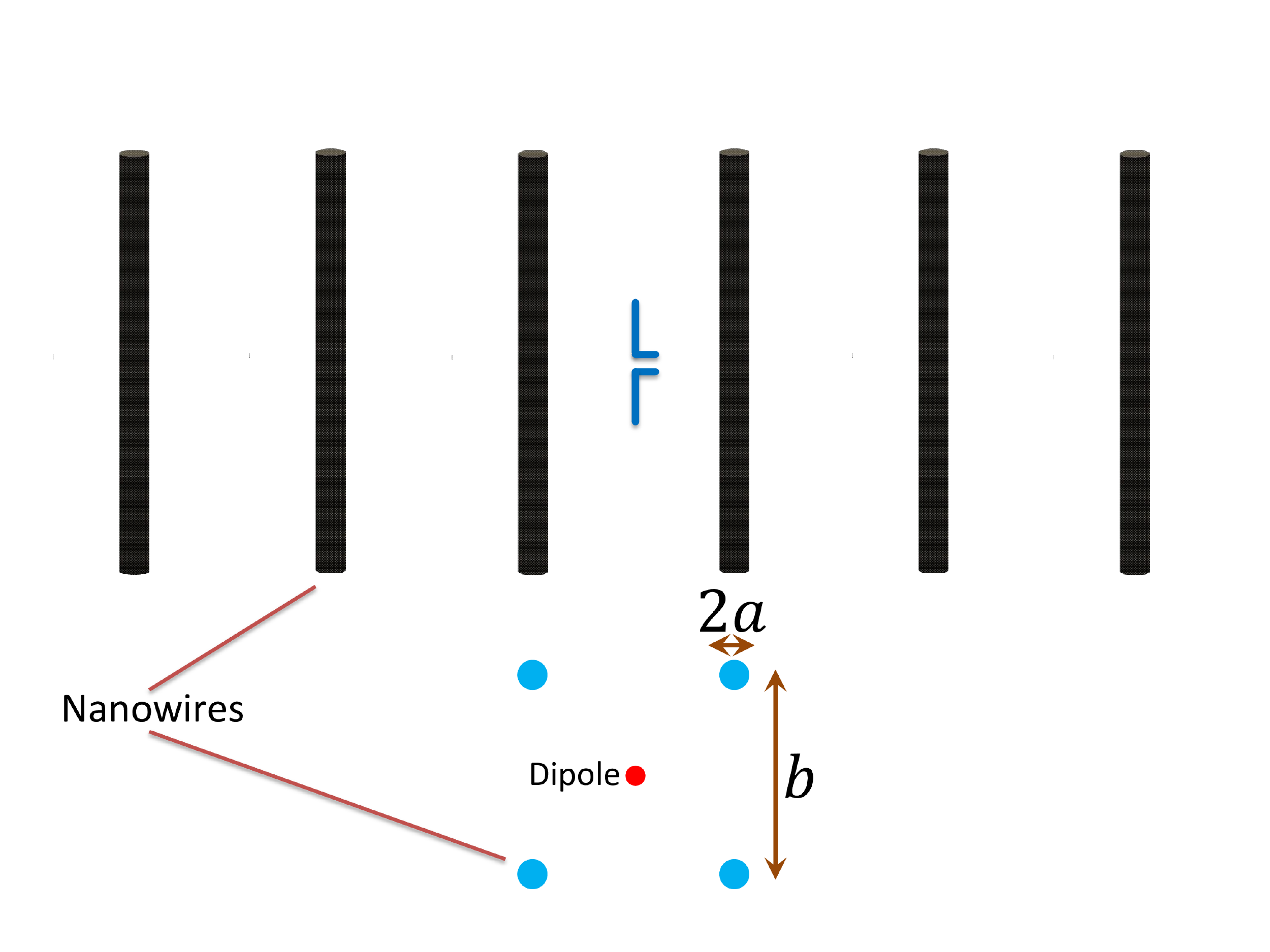}
\caption{Geometry and orientation of the sub-wavelength electric dipole source in the metamaterial sample.}
\label{fig:vertical}
\end{figure}

The geometry and orientation of the dipole is shown in Fig.~\ref{fig:vertical}.

\begin{figure}[t!]\centering
\subfigure[]{\includegraphics[width=7.5cm]{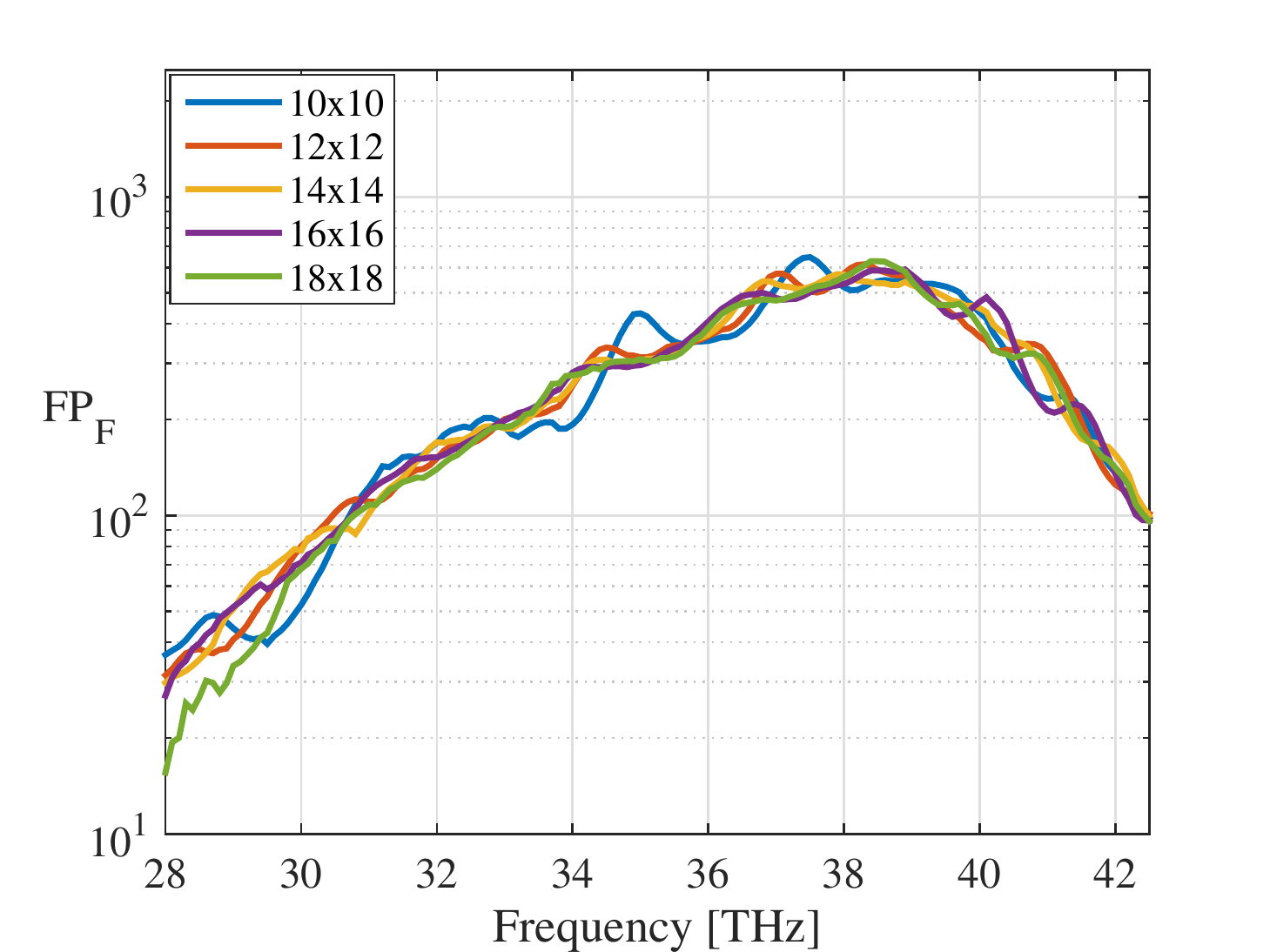}
\label{fig:PVZ}}
\subfigure[]{\includegraphics[width=7.5cm]{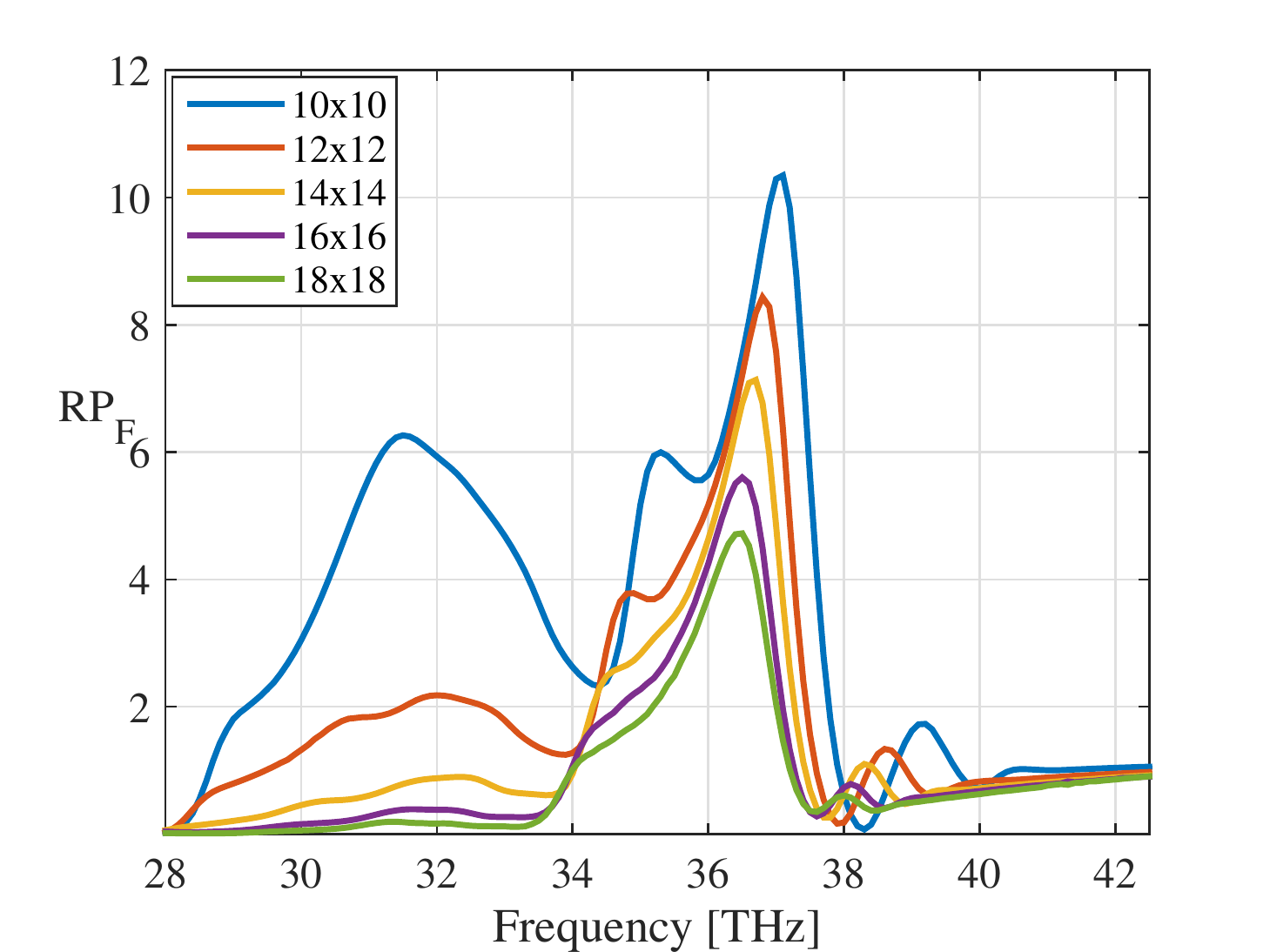}
\label{fig:PVR}}
\caption{Purcell factor for different sizes of lithium tantalate cubic wire medium. (a)--Full Purcell factor. (b)--Radiative Purcell factor. Here $a=32$ nm and $b=200$ nm.}
\end{figure}

Figure~\ref{fig:PVZ} depicts the full Purcell factor versus frequency. To be sure that our effect is not distorted by dimensional resonances, we simulated the cubic sample with five sizes of the edge, corresponding to $10\times10$, $12\times12$, $14\times14$, $16\times16$ and $18\times18$ nanowires, respectively (the internal structure of the wire medium was kept the same, that means the length of nanowires varying from 1.8 microns to 3.4 microns). As it is seen, the dimensional resonances do not enter the selected frequency range i.e. the full Purcell factor in all cases emulates the infinite wire medium. About the frequency $f=38$ THz, the full Purcell factor has its maximum value, whereas the homogenization model predicts the maximum at $f=39$ THz. This small error is not surprising. First,  the non-local homogenization model though most accurate of known effective-medium models of wire media is still an analytical approximation. Second, the whole theory of Sec.~\ref{sec:theory} neglects optical losses, though they are quite substantial. Notice, that these high losses make the resonance of the Purcell factor in Fig.~\ref{fig:PVZ} not very pronounced. Indeed, in a lossy medium the topological transition cannot be sharp. It happens not at a single frequency but across a certain frequency interval. Another reason why the resonance in Fig.~\ref{fig:PVZ} is not sufficiently pronounced is the logarithmic scale of the plot.

Whereas the full Purcell factor can exceed 600 at the transition frequency, the radiative Purcell factor shown in Fig.~\ref{fig:PVR} is not very high. However, it is still remarkable at the resonance. The global maximum was achieved for the sample of $10\times10$ nanowires, and the maximum occurs at nearly $f=37$ THz. If the size of the sample is smaller, the sample becomes mesoscopic (the full Purcell factor starts to feel the sample size), and on the other hand, if the size of the sample is larger, the maximum of radiative Purcell factor decreases monotonously versus the size which means that the internal reflections have no visible impact. At the resonant frequency, the reflection at the interface with free space is not significant because the wave impedance of the sample is relatively matched to the free-space wave impedance. This matching allows the electromagnetic wave to exit from the sample. The smallness of the radiative Purcell factor compared to the full one can result from the attenuation of the radiation in the sample.  
\begin{figure}[h!]\centering
\subfigure[]{\includegraphics[width=7.5cm]{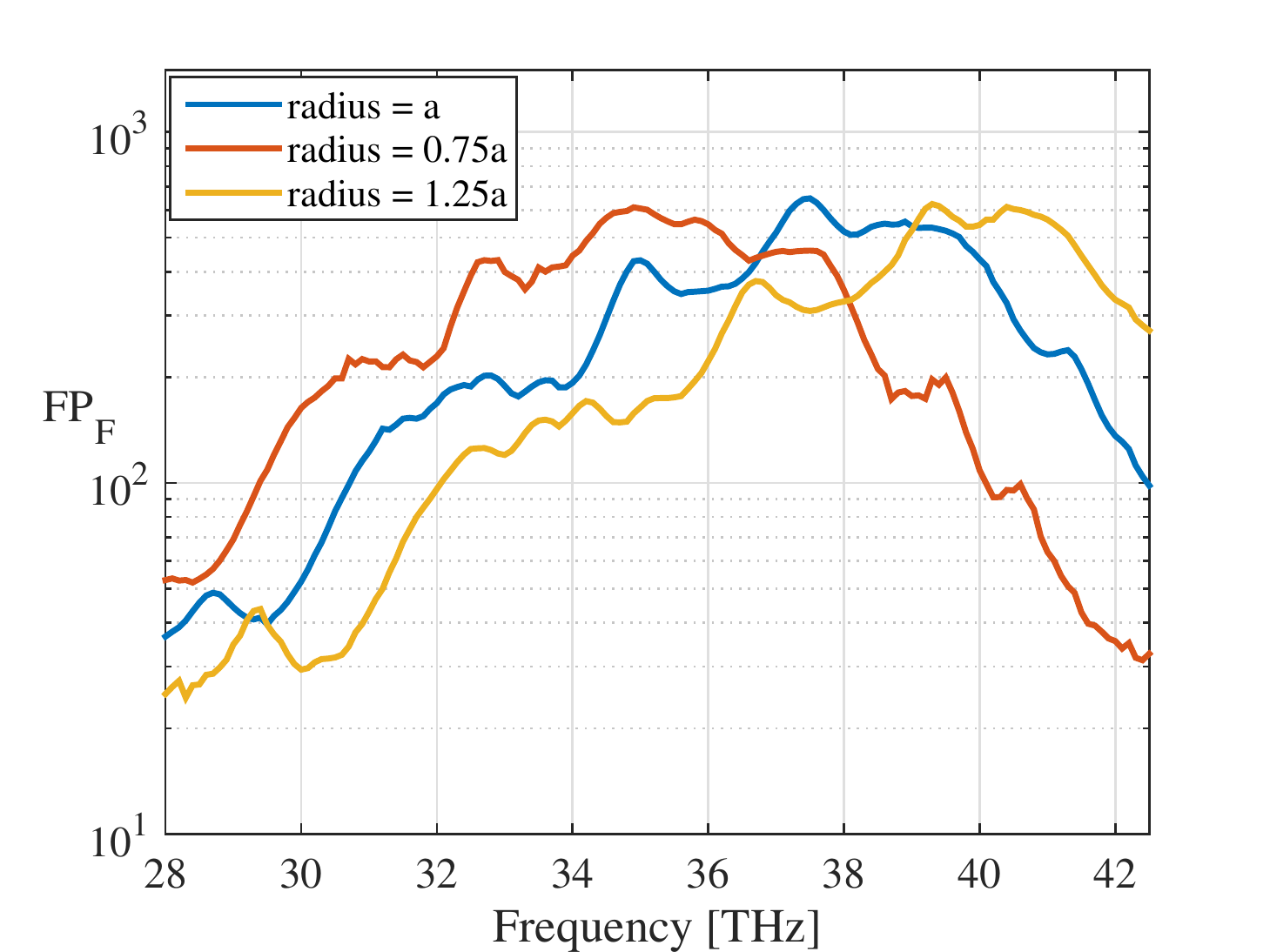}
\label{fig:PVZR}}
\subfigure[]{\includegraphics[width=7.5cm]{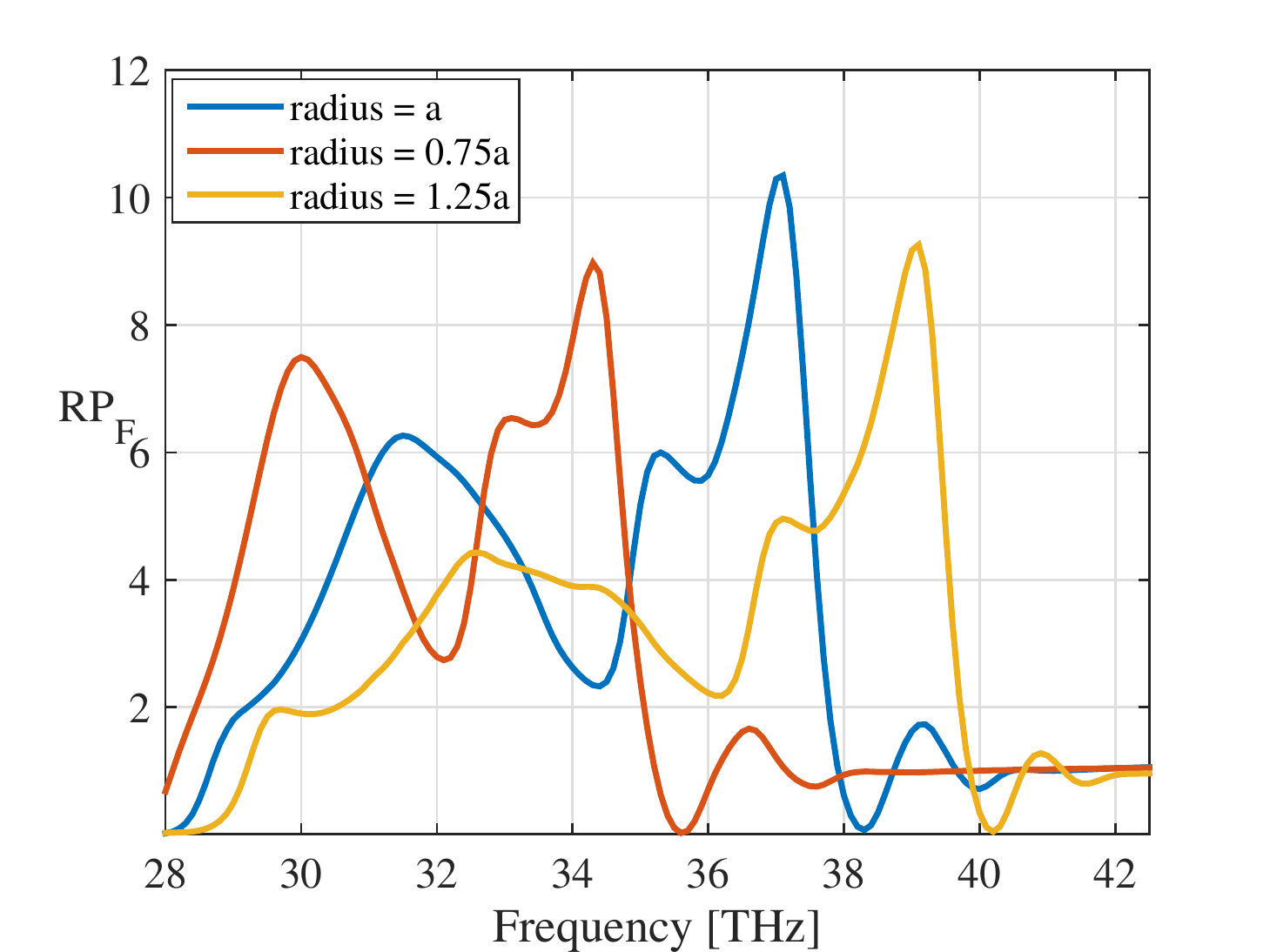}
\label{fig:PVRR}}
\caption{Purcell factor for different radii of wires in lithium tantalate cubic wire medium including $10\times10$ nanowires. (a)--Full Purcell factor. (b)--Radiative Purcell factor. Here $b=200$ nm.}
\end{figure}

As to other frequencies, around the topological phase transition range, the internal reflections become important due to the wave impedance mismatch. Close to the resonant frequency, the radiative Purcell factor drops dramatically because the metamaterial sample experiences the strong wave impedance mismatching at the interface. This mismatch definitely implies the standing waves, however, the dimensional resonances do not arise due to the high attenuation. At low frequencies where the open-surface regime  only exists, a local maximum is seen that is most visible for the sample of $10\times10$ nanowires. This local maximum at $f=32$ THz is also due to the impedance matching since it exists for all sizes of the sample.

Figures~\ref{fig:PVZR} and \ref{fig:PVRR} compare the full Purcell factor and radiative Purcell factor, respectively, for several radii of nanowires while the lattice constant is fixed. According to Eq.~\ref{eq:condition}, by changing the radius i.e. the volume fraction, we change $\varepsilon_{\rm{r}}$ and, consequently, shift the frequency of the topological transition. This frequency shift is clearly seen in the figures for both full and radiative Purcell factors.

\subsection{Transversal dipole emitter}
In this subsection, we report the results of similar calculations for the transverse dipole (with the same absolute value of the dipole moment as above) and compare to the previous ones obtained for an axial one. For better clarity of our plots, we show the results only for two cubic samples -- $10\times10$ and $12\times12$ nanowires.
\begin{figure}[h!]\centering
\subfigure[]{\includegraphics[width=7.5cm]{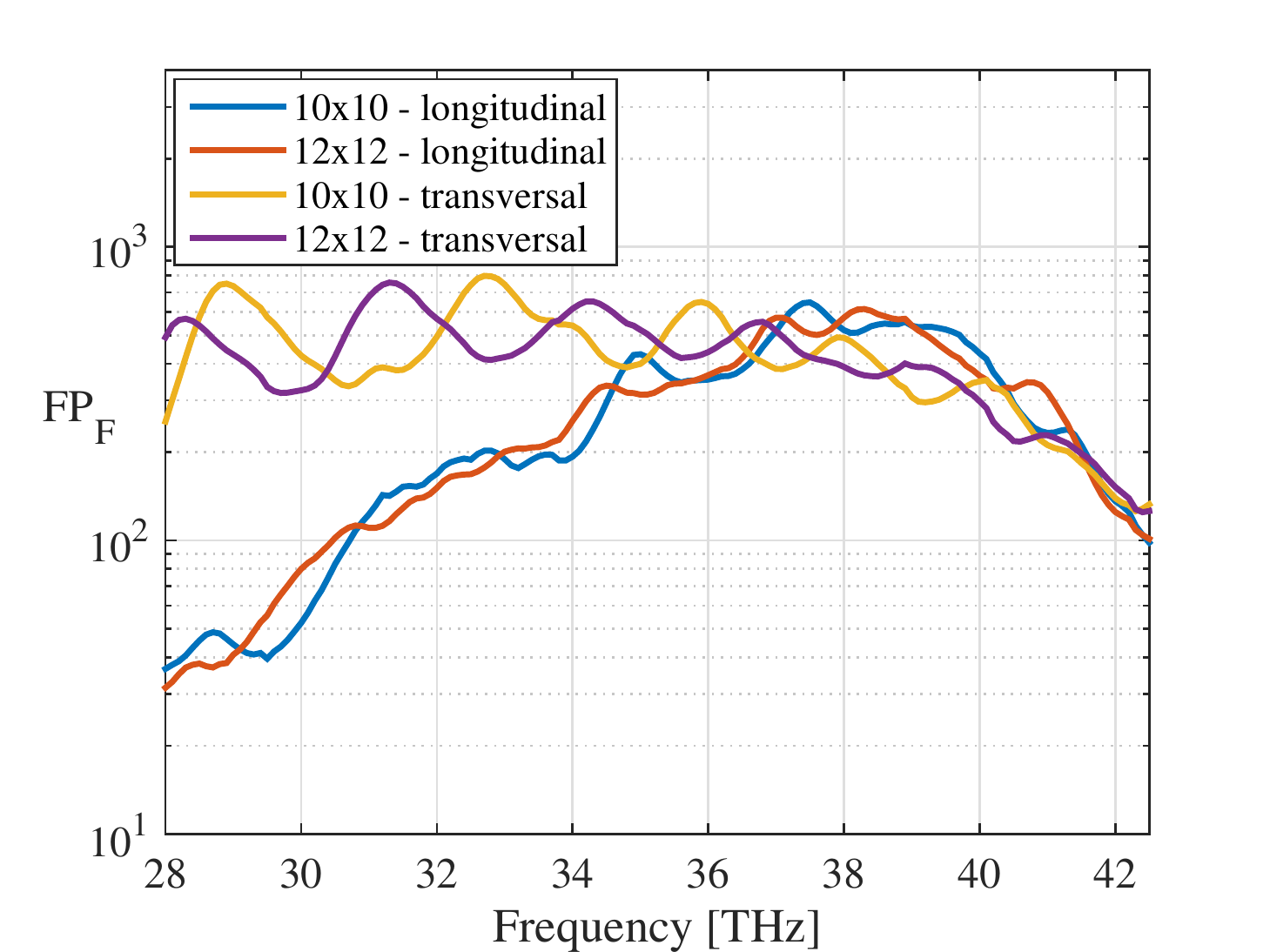}
\label{fig:PHZ}}
\subfigure[]{\includegraphics[width=7.5cm]{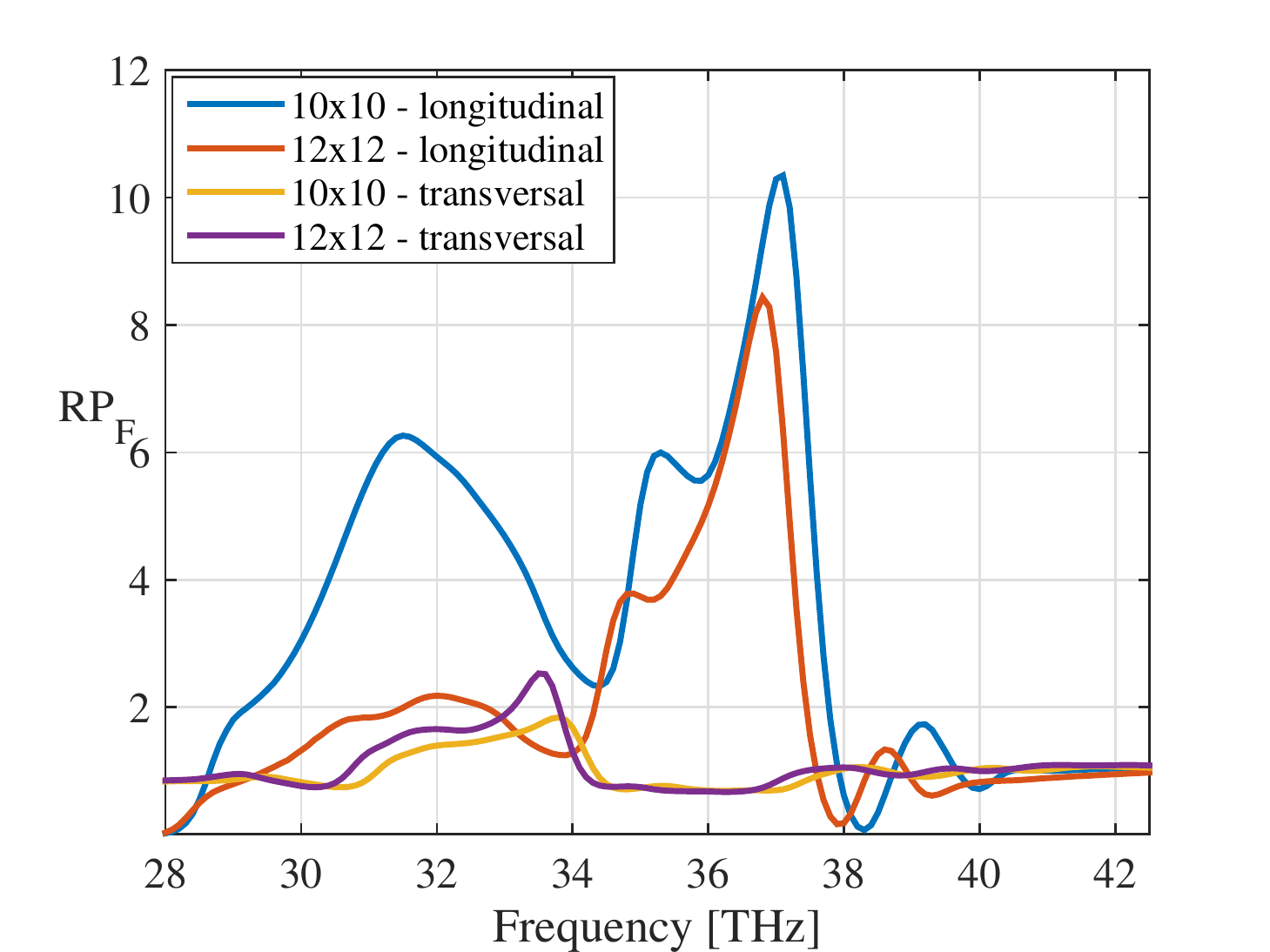}
\label{fig:PHR}}
\caption{Purcell factor for two different orientations of the dipole. Two samples of lithium tantalate cubic wire medium including $10\times10$ and $12\times12$ nanowires are considered. (a)--Full Purcell factor and (b)--radiative Purcell factor. Here, $a=32$ nm and $b=200$ nm.}
\end{figure}

Figure~\ref{fig:PHZ} shows that the transverse dipole does not feel the topological transition. Definitely, the radiation decreases above $f=40$ THz, where the dispersion surface is closed. However, at the frquency range where the topological transition happens for the axial dipole, the Purcell factor for the transverse one is the same as below -- in the \textquotedblleft{standard}\textquotedblright open-surface regime. This fully agrees with our theoretical expectations. The Purcell factor of the transverse dipole turns out to be smaller than that of the axial dipole (this result keeps the same for other sizes of the sample) at the transition. Especially, the insensitivity of the transverse dipole to the topological transition is seen in the plot for the radiative Purcell factor depicted in Figure~\ref{fig:PHR}. The longitudinal (axial) dipole is much more efficient for radiation to free space than the transversal one.
For it, the radiative Purcell factor does not exceed 2 and equals unity in the range of the topological transition and around. Definitely, this is the result of the strong wave impedance mismatch. Since the material is anisotropic, the regime $\varepsilon_\parallel\approx 0$ mimics the $\varepsilon$-near-zero regime only for the radiation of the axial dipole (with dominating axial polarization), whereas the radiation of the transverse dipole almost does not feel the zero of $\varepsilon_\parallel$  and turns out to be confined inside the sample. Therefore the contrast between $RP_{\rm{F}}$ of the axial and transverse dipoles at about the transition frequency is 10-fold.


\section{Conclusions}
\label{sec:conclusion}
In this work, we have theoretically revealed and studied topological phase transition in wire medium which may occur in the frequency range where the homogenization model is still valid. We utilized the spatial dispersion inherent to wire medium in order to engineer the unbounded spatial spectrum of propagating eigenmodes at the transition frequency.
The open and closed parts of the dispersion surface in this regime touch each other at a special point in the reciprocal space of the lattice.
At this point the real part of the axial component of the effective permittivity is zero. We found the practical design parameters for lithium tantalate nanowires which allow to implement this regime.

Due to this effect (topological phase transition over the wavevector axis at one certain frequency), we obtained by exact numerical simulations a dramatic increase of the radiated power of a sub-wavelength electric dipole located at the center of the finite-size sample of our wire medium. The highest Purcell factor corresponds to the axially oriented dipole, that represents the contrast to previously known results where the high Purcell factors of wire media were reported only for the transverse dipoles.  Whereas the radiation of the axial dipole is resonant with the maximum at the transition frequency, the radiation of the transverse dipole is rather stable in the
frequency range of open-surface dispersion and smoothly decreases at higher frequencies where the dispersion surface becomes closed.
We also investigated the radiative Purcell factor. It attains one order of magnitude for an axial dipole.

Finally, since lithium tantalate is a lossy material, and the radiative Purcell factor we have reported is modest, we plan to reveal the same regime for wire medium of SiC or GaN, where
$RP_{\rm{F}}$ may approach more closely to the resonant value of the full Purcell factor e.g. attain two orders of magnitude. Then we will organize an experimental verification of our effect.




\begin{thebibliography}{00}


\bibitem{veselago}
V.~G.~Veselago,
The electrodynamics of substances with simultaneously negative values of $\varepsilon$ and $\mu$,
Sov.~Phys.~Usp.~\textbf{10}, 509--514 (1968).

\bibitem{pendry}
J.~B.~Pendry,
Optical conformal mapping,
Science~\textbf{312}, 1777--1780 (2006).

\bibitem{leonhardt}
U.~Leonhardt,
Negative refraction makes a perfect lens,
Phys.~Rev.~Lett.~\textbf{85}, 3966--3969 (2000).

\bibitem{kivshar}
A.~Poddubny, I.~ Irosh, P.~Belov and Y.~Kivshar,
Hyperbolic metamaterials,
Nature Photonics~\textbf{7}, 948--957 (2013).

\bibitem{smith}
D.~R.~Smith and D.~Schurig,
Electromagnetic wave propagation in media with indefinite permittivity and permeability tensors,
Phys.~Rev.~Lett.~\textbf{90}, 077405 (2003).

\bibitem{jacob}
Z.~Jacob, I.~I.~Smolyaninov and E.~E.~Narimanov,
Broadband Purcell effect: radiative decay engineering with metamaterials,
Appl.~Phys.~Lett.~\textbf{100}, 181105 (2012).

\bibitem{belov}
A.~N.~Poddubny, P.~A.~Belov and Y.~S.~Kivshar,
Spontaneous radiation of a finite-size dipole emitter in hyperbolic media,
Phys.~Rev.~A~\textbf{84}, 023807 (2011).

\bibitem{yan}
W.~Yan, M.~Wubs and N.~A.~Mortensen,
Hyperbolic metamaterials: nonlocal response regularizes broadband supersingularity,
Phys.~Rev.~B~\textbf{86}, 205429 (2012).

\bibitem{purcell1}
E.~M.~Purcell,
Spontaneous emission probabilities at radio frequencies,
Phys.~Rev.~\textbf{69}, 681--689 (1946).

\bibitem{purcell2}
C.~Sauvan, J.~P.~Hugonin, I.~S.~Maksymov and P.~Lalanne,
Theory of the spontaneous optical emission of nanosize photonic and plasmon resonators,
Phys.~Rev.~Lett.~\textbf{110}, 237401 (2013).

\bibitem{purcell}
L.~Novotny and B.~Hecht, \emph{Principles of Nano-Optics} (Cambridge University Press, Cambridge, UK, 2006).

\bibitem{constantin}
C.~R.~Simovski, P.~A.~Belov, A.~V.~Atrashchenko and Y.~S.~Kivshar,
Wire metamaterials: physics and applications,
Adv.~Mater.~\textbf{24}, 4229--4248 (2012).

\bibitem{chebykin}
A.~V.~Chebykin, A.~A.~Orlov, C.~R.~Simovski, Yu.~S.~Kivshar and P.~A.~Belov, 
Nonlocal effective parameters of multilayered metal-dielectric metamaterials, 
Phys.~Rev.~B~\textbf{86}, 115420 (2012).

\bibitem{mario}
M.~G.~Silveirinha,
Nonlocal homogenization model for a periodic array of $\epsilon$-negative rods,
Phys.~Rev.~E~\textbf{73}, 046612 (2006).

\bibitem{lifshitz}
I.~M.~Lifshitz,
Anomalies of electron characteristics of a metal in the high pressure region,
Sov.~Phys.~JETP~\textbf{11}, 1130--1135 (1960).

\bibitem{krishnamoorthy}
H.~N.~S.~Krishnamoorthy, Z.~Jacob, E.~Narimanov, I.~ Kretzschmar and V.~M.~Menon,
Topological transitions in metamaterials,
Science~\textbf{336}, 205--209 (2012).

\bibitem{yang}
X.~Yang, C.~Hu, H.~Deng, D.~Rosenmann, D.~A.~Czaplewski and J.~Gao,
Experimental demonstration of near-infrared epsilon-near-zero multilayer metamaterial slabs,
Opt.~Express~\textbf{21}, 23631--23639 (2013).

\bibitem{gomez}
E.~Reyes-Gomez, S.~B.~Cavalcanti, L.~E.~Oliveira and C.~A.~A.~de Carvalho,
Metric-signature topological transitions in dispersive metamaterials,
Phys.~Rev.~E~\textbf{89}, 033202 (2014).

\bibitem{molesky}
S.~Molesky, C.~J.~Dewalt, and Z.~Jacob,
High temperature epsilon-near-zero and epsilon-near-pole metamaterial emitters for thermophotovoltaics,
Optics Express~\textbf{21}, A96--A110 (2013).

\bibitem{kabashin}
A.~V.~Kabashin, P.~Evans, S.~Pastkovsky, W.~Hendren, G.~A.~Wurtz, R.~Atkinson, R.~Pollard, V.~A.~Podolskiy and A.~V.~Zayats, 
Plasmonic nanorod metamaterials for biosensing, 
Nature Mater.~\textbf{8}, 867--871 (2009).

\bibitem{poddubny}
A.~N.~Poddubny, P.~A.~Belov and Y.~S.~Kivshar,
Purcell effect in wire metamaterials,
Phys.~Rev.~B.~\textbf{87}, 035136 (2013).

\bibitem{vorobev}
V.~V.~Vorobev and A.~V.~Tyukhtin, 
Nondivergent Cherenkov radiation in a wire metamaterial, 
Phys.~Rev.~Lett.~\textbf{108}, 184801 (2012).

\bibitem{marcuvitz}
L.~B.~Felsen and N.~Marcuvitz, \emph{Radiation and Scattering of Waves} (IEEE Press, New York, 1994).

\bibitem{gorlach}
M.~A.~Gorlach and P.~A.~Belov,
Effect of spatial dispersion on the topological transition in metamaterials,
Phys.~Rev.~B~\textbf{90}, 115136 (2014).

\bibitem{polaritonics}
Mahi R.~Singh, 
Polaritonics in nanowires made from dispersive materials, 
Phys.~Rev.~B~\textbf{80}, 195303 (2009).

\bibitem{kittel}
C.~Kittel, \emph{Introduction to Solid State Physics} (Wiley, New York, 1976).

\bibitem{schall}
M.~Schall, H.~Helm and S.~R.~Keiding,
Far infrared properties of electro-optic crystals measured by THz time-domain spectroscopy,
Int.~J.~Infrared Millim.~Waves~\textbf{20}, 595--604 (1999).

\bibitem{crimmins}
T.~F.~Crimmins, N.~S.~Stoyanov and K.~A.~Nelson,
Heterodyned impulsive stimulated Raman scattering of phonon–polaritons in LiTaO3 and LiNbO3,
J.~Chem.~Phys.~\textbf{117}, 2882--2896 (2002).


\end{thebibliography}
\end{document}